\documentclass[10pt,times,mathptm,onecolumn,conference]{IEEEtran}

\usepackage{hyperref}
\usepackage{multirow}
\usepackage[]{graphicx}
\usepackage{color}
\usepackage{algorithm}
\usepackage{algorithmic}
\usepackage[]{amsmath,theorem}
\usepackage[]{amssymb}
\usepackage[]{epsfig}
\usepackage{enumerate}
\usepackage{color}
\usepackage{wrapfig}
\usepackage{subfig}
\usepackage{qcircuit}
\newtheorem{theorem}{Theorem}
\newtheorem{definition}{Definition}
\newtheorem{corollary}{Corollary}
\newtheorem{example}{Example}
\begin{document}

\title{Minimization of Quantum Circuits using Quantum Operator Forms}
\author{\IEEEauthorblockN{Martin Lukac\IEEEauthorrefmark{1}, Michitaka Kameyama\IEEEauthorrefmark{1},Marek Perkowski\IEEEauthorrefmark{2},Pawel Kerntopf\IEEEauthorrefmark{3}}
\IEEEauthorblockA{\IEEEauthorrefmark{1}Graduate School of Information Sciences, Tohoku University, Sendai, Japan\\ Email: \{lukacm,kameyama\}@ecei.tohoku.ac.jp}
\IEEEauthorblockA{\IEEEauthorrefmark{2}Department of Electrical and Computer Engineering, Portland State University, Portland, OR, USA\\ Email: mperkows@pdx.edu}
\IEEEauthorblockA{\IEEEauthorrefmark{3}Institute of Computer Science, Warsaw University of Technology, Warsaw, Poland \\
 Department of Theoretical Physics and Computer Science, University of Lodz, Lodz, Poland, Email: p.kerntopf@ii.pw.edu.pl}
}

\maketitle

\begin{abstract}
In this paper we present a method for minimizing reversible quantum circuits using the  Quantum Operator Form (QOF); a new representation of quantum circuit and of quantum-realized reversible circuits based on the CNOT, CV and CV$^\dagger$ quantum gates. The proposed form is a quantum extension to the well known Reed-Muller but unlike the Reed-Muller form, the QOF allows the usage of different quantum gates. Therefore QOF permits minimization of quantum circuits by using properties of different gates than only the multi-control Toffoli gates. We introduce a set of minimization rules and a pseudo-algorithm that can be used to design circuits with the CNOT, CV and CV$^\dagger$ quantum gates. We show how the QOF can be used to minimize reversible quantum circuits and how the rules allow to obtain exact realizations using the above mentioned quantum gates. 
\end{abstract}

\section{Introduction}

In quantum and reversible circuits synthesis methods various representations are used for minimization, mapping or manipulation. The most famous of these forms is the Reed-Muller family (also known as Zhegalkin Polynomials) of expansions~\cite{zhegalkin:29,muller:54,reed:54}. Reed-Muller form is particularly well suited for reversible logic because it is based on two-level AND/EXOR which can be directly mapped into reversible circuit using the Toffoli gates for instance. 

However, it is now well known that mapping reversible circuits to reversible gates and then performing a technology mapping - such as mapping to the set of elementary gates CNOT, CV/CV$^\dagger$ or to a LNN~\cite{chakrabarti:07, chakrabarti:07a, hirata:09, saeedi:10, perkowski:11, matsuo:11} architecture restricted set does not always generate a minimal result. Also, it is not possible to obtain minimal realization on the level of quantum gates when minimizing solely on the level of Toffoli gates~\cite{sasanian:12}. 

In this paper we propose the so called Quantum Operator Form (QOF) which is a quantum-expanded counter part of the classical Reed-Muller expansion. Starting from a set of simple rules extracted from the interaction of the CV and CV$^\dagger$ operators we generalize them to various conditions and provide a mechanism to map arbitrary reversible circuit directly to quantum primitives. Using the rules of the interaction between quantum operators we then show how QOF permits for minimization and how to obtain forms of the QOF.

This paper is organized as follows. Section~\ref{sec:brmf} introduces the Reed-Muller and more general the ESOP forms used for representation of reversible circuits. Section~\ref{sec:pnqo} describes the operators used and their interaction based. Section~\ref{sec:notation} introduces a notation that allows to express operator required to allow a general quantum circuit expressions. Section~\ref{sec:qocf} explains the principle of the QOF and Section~\ref{sec:minimization} provides an example demonstrating the capabilities of the QOF and simple rules of minimization. Finally Section~\ref{sec:canonicity} introduces the notion of the weak canonicity of QOFs and a conclusion concludes the paper in Section~\ref{sec:conclusion}.

\section{Reed-Muller Forms}
\label{sec:brmf}
In reversible logic the Reed-Muller (RM) form is very popular because a logic function expressed in the RM form can be directly mapped to a sequence of Toffoli gates. The RM form can be expressed by
\begin{equation}
F = \bigoplus_{i=1}^{2^n} \alpha_i c_i
\end{equation} 
with $\alpha_i$ being Boolean coefficients indicating if the $c_i$ term is part of the RM expression or not~\cite{sasao:99}. The $c_i$ term represents any product of positive or negative polarity literals. Consequently depending on the type of the term used there are three main classes of the RM forms: the PPRM (positive polarity Reed-Muller), FPRM (fixed polarity Reed-Muller) and GRM (generalized Reed-Muller). Examples of these forms are given in eq.~\ref{eq:rmexample}
\begin{equation}
\begin{split}
F_{PPRM} =& abc\oplus ad\oplus bce\oplus ade\\
F_{FPRM} =& abc\oplus  a\bar{d}e\oplus bc\bar{d}\\
F_{GRM}  =& abc\oplus \bar{a}bd\oplus e\bar{d}
\end{split}
\label{eq:rmexample}
\end{equation}

The difference between these forms and ESOP is that ESOP in general can be minimized by combining minterms to obtain one RM expression. For instance let $f_{esop} = abc\oplus a\bar{b}c \oplus abd$. Naturally, the two product terms $abc$ and $a\bar{b}c$ in $f$ result in $ac$. 

\begin{figure}[bht]
\centering
\includegraphics[width=0.3\textwidth]{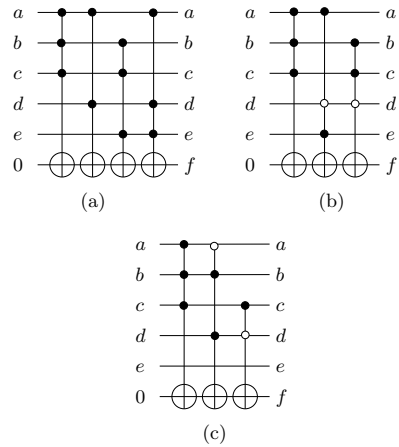}
\caption{\label{fig:rmexample} Example circuits for (a) PPRM, (b) FPRM, (c) GRM.}
\end{figure}

The RM forms are very well suited for reversible circuit design because each term is directly mapped to a Toffoli gate. For instance the three forms described in eq.~\ref{eq:rmexample} are shown in corresponding circuits in Figure~\ref{fig:rmexample}.

\begin{definition}[Toffoli gate]
A Toffoli gate is a single NOT operator controlled by a product of literals with positive and/or negative polarities.
\end{definition}
A literal is a variable or a negation of a variable and thus a Toffoli is a quantum gate that can have positive or negative controls. In pictorial representation of reversible gates positive controls are denoted by "black dots" and negative controls are denoted by "white dots" (see Figure~\ref{fig:rmexample}).

\begin{corollary}[Toffoli gate decomposition]
A Toffoli gate with $n>2$ control bits can always be decomposed to $2^{n-2}+1$ Toffoli gates with $2$ control bits and with $n-2$ ancilla bits.
\label{def:cdecomposition}
\end{corollary}

Notice that corollary~\ref{def:cdecomposition} is quite easy to prove and thus in the case if the ancilla bits are not being reused the number of Toffoli gates in the decomposition is halved. 

\begin{figure}[bht]
\centering
\includegraphics[width=0.25\textwidth]{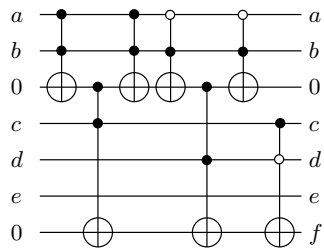}
\caption{\label{fig:tofdecomposition} Circuit from Figure~\ref{fig:rmexample}(c) expanded to Toffoli gates with two control lines.}
\end{figure}

\begin{corollary}[Upper bound on the Reed-Muller Form I]
An arbitrary Boolean function expressed in the RM form with $k>1$ product terms can be built with a maximum of $2^{n_k-2}+k; n > 2$ two-bit controlled Toffoli gates; with $n_k$ being the number of control bits in the k$^{th}$ product term.
\end{corollary}

The proof is easy to be verified by simply applying to arbitrary RM form the decomposition from corollary~\ref{def:cdecomposition}.

\section{Permutative and Non-permutative Quantum Operators}
\label{sec:pnqo}

\begin{figure}[bht]
\centering 
	\subfloat[]{
\Qcircuit @C=1em @R=.7em {
	\lstick{a} & \ctrl{1} & \qw & \ctrlo{1} \qw\\
	\lstick{b} & \gate{V/V^\dagger} & \qw & \gate{V/V^\dagger} \qw
}}
\hspace{1.5em}
	\subfloat[]{\Qcircuit @C=1em @R=.9em {
     \lstick{a} & \ctrl{1} & \qw & \ctrlo{1} &\qw\\
     \lstick{b} & \targ & \qw & \targ &\qw
	}}
\caption{\label{fig:allgates}The two-qubit primitive gates used in QOF (a) the CV/CV$^\dagger$ gate with positive and negative control and (b) the CNOT gate with positive and negative control bit.}
\end{figure}
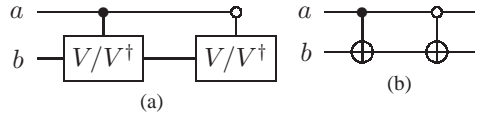

The proposed Quantum Operator Form in this paper is using the CV/CV$^\dagger$ and CNOT two-qubit quantum gates. Moreover the quantum gates are used with both positive and negative control variables. All the used gates are shown in Figure~\ref{fig:allgates}.

%

\begin{figure}[bht]
	\centering
\includegraphics[width=0.5\textwidth]{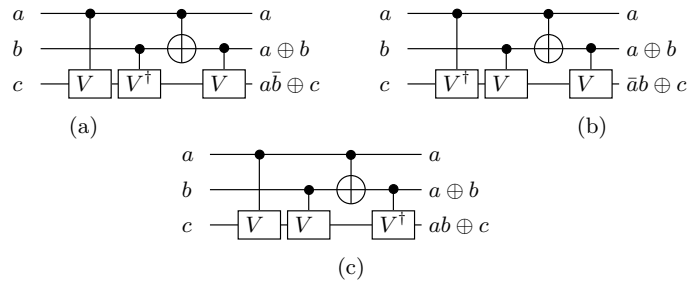}
\caption{\label{fig:peres}Three types of the Peres gates generated by the permutation of the V and V$^\dagger$ operators.}
\end{figure}

The V and V$^\dagger$ gates are known for being the so called square-root of the NOT gate. Equation~\ref{eq:prop1} represents this property in a formal way.
\begin{equation}
\begin{split}
V*V &= V^\dagger * V^\dagger = NOT\\
V*V^\dagger &= V^\dagger * V = I\\
V*NOT &= NOT*V = V^\dagger\\
V^\dagger*NOT &= NOT*V^\dagger = V
\end{split}
\label{eq:prop1}
\end{equation}
The V quantum operator has its matrix shown in eq.~\ref{eq:vmatrix} and V$^\dagger$ is its hermitian conjugate. 
\begin{equation}
\begin{split}
V = \begin{pmatrix}\frac{1+i}{2}& \frac{1-i}{2}\\\frac{1-i}{2}& \frac{1+i}{2}\end{pmatrix} &\;\; V^\dagger = \begin{pmatrix}\frac{1-i}{2}& \frac{1+i}{2}\\\frac{1+i}{2}& \frac{1-i}{2}\end{pmatrix}
\end{split}
\label{eq:vmatrix}
\end{equation}
Consequently, the CV and the CV$^\dagger$ quantum gates are also called root squares of the CNOT permutative gate. The property from eq.~\ref{eq:prop1} is known to have given rise to a family of Peres gates, all at the same cost~\cite{lukac:05a}. Some of them are shown in Figure~\ref{fig:peres}.

\begin{figure}[bht]
\centering
\subfloat[$ab\oplus c$ decomposed using CNOT/CV/CV$^\dagger$ and with CV/CV$^\dagger$]{\label{fig:toffolia}\includegraphics[width=0.45\textwidth]{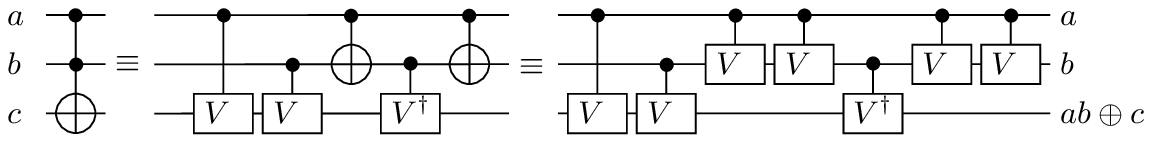}}
\subfloat[$\bar{a}b\oplus c$ decomposed using CNOT/CV/CV$^\dagger$ and with CV/CV$^\dagger$]{\label{fig:toffolib}\includegraphics[width=0.45\textwidth]{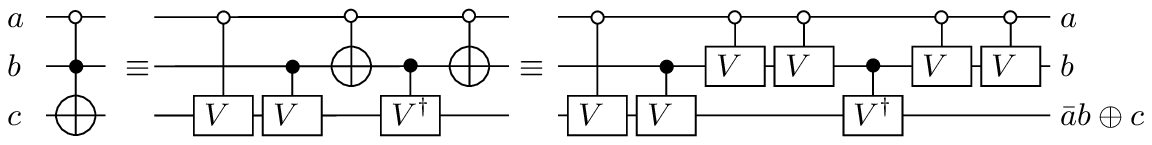}}
\caption{Example of Quantum Operator Decomposition of the Toffoli gate.}
\end{figure}

\begin{corollary}[Quantum Operator Decomposition]
A two-qubit controlled NOT gate (Toffoli gate) can be decomposed into five two qubit quantum gates that can be directly mapped into hermitian operators such as electromagnetic or laser pulses. Examples of such decompositions are shown in Figure~\ref{fig:toffolia}. The decomposition can be further extended to seven quantum CV/CV$^\dagger$ two-qubit quantum operators. Examples of such decompositions are shown in Figure~\ref{fig:toffolib}.
\label{def:qdecomposition}
\end{corollary}

Corollary~\ref{fig:toffolib} is based on the well known decomposition of Toffoli gates using CV/CV$^\dagger$ primitives and thus it is easy to notice that it is true.

\begin{corollary}[Upper bound on the Reed-Muller Form II]
An arbitrary Boolean function expressed in the RM form with $k>1$ product terms can be built with a maximum of $5*2^{n_k-2}+k; n > 2$ two-qubit quantum operators.
\label{def:quantumbound}
\end{corollary}

Again corollary~\ref{def:quantumbound} is easily provable by applying the decomposition from corollary~\ref{def:qdecomposition} to a RM form. 

\section{Notation for Generalized Reed-Muller Forms}
\label{sec:notation}


To represent the operator expansions as described in Corollary~\ref{def:cdecomposition} and~\ref{def:qdecomposition} an extended  notation is used. This notation allows to express not only product terms controlling a NOT gate but also product terms controlling an arbitrary unitary operator as well as additional information allowing to precisely locate a particular quantum operator one some variables. 
To express two-qubit operator such as Controlled-V we use the following notation:
\begin{equation}
a^V_{t} \equiv \;\;\;\;\Qcircuit @C=1em @R=.7em {
     \lstick{a} & \ctrl{1} & \qw \\
      \lstick{t} & \gate{V} & \qw 
}
\label{eq:singleop}
\end{equation}
 where the literal or product of literals are the control variables, the subscript shows the target variable of the applied unitary transform and the superscript shows what unitary transform is being applied. Eq.~\ref{eq:singleop} shows an example of this generalized notation. A multi-qubit controlled terms use the following representation:
\begin{equation}
(abd)^V_t \equiv abd^V_t \equiv \;\;\;\;\Qcircuit @C=1em @R=.9em {
     \lstick{a} & \ctrl{1} & \qw \\
     \lstick{b} & \ctrl{2} & \qw \\
     \lstick{c} & \qw & \qw \\
     \lstick{d} & \ctrl{1} & \qw \\
      \lstick{t} & \gate{V} & \qw 
}
\label{eq:multiop}
\end{equation}
Eq.~\ref{eq:singleop} and ~\ref{eq:multiop} show that anytime a variable is present in a product term, it always controls an operator that changes the value of a target bit(s). Finally note that a multi-qubit function can be such that generates output on some of the input qubits and does not use dedicated output qubits: e.g. CNOT or SWAP. In such case the above introduced notation using the dedicated output variable $t$ is modified to the target variable $b$ by changing the subscript:
\begin{equation}
\begin{split}
a\oplus b =& (a\oplus b)_b^{NOT} \\ =& (a\oplus b)_b =\;\;\;\;\Qcircuit @C=1em @R=.9em {
     \lstick{a} & \ctrl{1} &\rstick{a} \qw\\
     \lstick{b} & \targ &\rstick{a\oplus b} \qw \\
}
\end{split}
\label{eq:cnotop}
\end{equation}

For the simplicity of notation and better understanding, the NOT operation is dropped from the superscript: This is shown by the equivalence of expressions in eq.~\ref{eq:cnotop}.

Notice that this notation creates cascades of either classical Toffoli gates or cascades of general quantum gates. For instance expanding the Toffoli gate a cascade of $CNOT,CV^\dagger,CNOT$ is created in such manner that the $cV^\dagger$ depends on the $CNOT$ gate. The dependency between gates in a cascade can be easily indicated by the correct subscripts and control variables, i.e. $(a\oplus b)_bb^\dagger(a\oplus b)$. However it is also required to point out that the order of the gates cannot be changed in the current form, i.e. $b^\dagger(a\oplus b)_b(a\oplus b) = b^\dagger$. Thus we introduce the $\circ$ operation as a notation for a sequence of gates that cannot be altered and that depend on each other with some intermediary variables. Using this notation a Toffoli gate written in the generalized notation and decomposed to its CV/CV$^\dagger$ and CNOT primitives will take the following form:
\begin{equation}
\begin{split}
F = & ab\oplus c\\
= &a^V_cb^V_c(a\oplus b)_b\circ b^{V\dagger}_c\circ (a\oplus b)_b
\end{split}
\end{equation}

Note that terms and operators connected by $\circ$ are considered as one and without manipulation cannot be separated.

Finally in the generalized RM notation for quantum circuit, the joining operation is not always the $XOR$. Thus individual terms are separated by the following rules:
\begin{itemize}
\item $\circ$ separates terms as described above,
\item a superscript indicating what operation is being applied to the last variable in the control term (example in Figure~\ref{eq:cnotop})
\item a superscript and a subscript indicating what unitary operator is being applied to which variable.
\end{itemize}

\section{Quantum Operator Forms}
\label{sec:qocf}
The Quantum Operator Form allows to obtain an optimized representation of the quantum circuit. However, the transformations leading to the final form may seem counterintuitive because they create more complex circuits and require gates that might not be realizable. However, these transformations formalize the beginning of a step by step effective optimization process. 
 
First, let us define the arbitrary controlled unitary operator:
\begin{definition}
Arbitrary controlled Unitary operator (ACUO) is a single qubit quantum operator that is controlled by positive and/or negative literals. 
\end{definition}
An example of ACUO using the NOT operator are shown in Figure~\ref{fig:rmexample}.

\begin{definition}[Uninterrupted Quantum Line]
A line segment in a quantum circuit between two control inputs \em a \em and \em b \em laying on this line is called uninterrupted if the two points can be put as close to each other as possible.
\label{def:uql}
\begin{figure}
\centering
\includegraphics[width=0.3\textwidth]{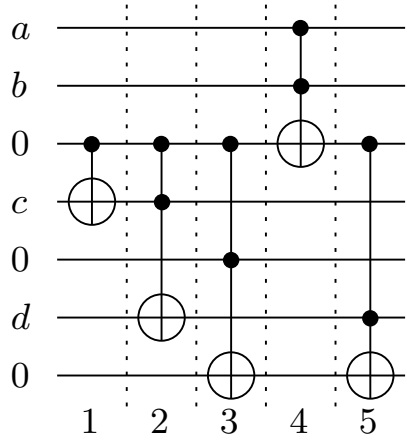}
\caption{\label{fig:interexample}Example of uninterrupted and interrupted lines: gates 1,2,3 and 5 are on a interrupted quantum line by gate 4. Gates 1 and 3 and 2 and 3 are on uninterrupted quantum lines but gates 1 and 2 are on interrupted quantum lines.}
\end{figure}
\end{definition}

\begin{corollary}
When the quantum gates having control inputs \em a \em and \em b \em are hermitian then the order of these gates can be reversed.
\end{corollary}

\begin{definition}[Interrupt Point]
An interrupt point is the output qubit of a quantum gate such as V, CNOT, Toffoli or so. 
\end{definition}

\begin{definition}[Terminal Gate]
A gate is called terminal if it has the target bit on one of the function output variables.
\end{definition}

\begin{figure}[bht]
\centering
\includegraphics[width=0.3\textwidth]{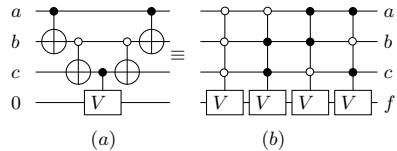}
\caption{\label{fig:nonterminaltransf}Expansions - Non-Terminal Gate Transformation - Illustrated on the $(\bar{a}\oplus b\oplus c)^V$.}
\end{figure}

\begin{corollary}[Non-Terminal Gate CNOT Transformation]
Any terminal CV/CV$^\dagger$ operator combined with a cascade of non-terminal quantum gates $CNOT$ can be expanded into a set of multi-qubit controlled terminal gates. Each of the resulting terminal gates corresponds to one of the terms of the expanded CNOT gate cascade. An example of such expansion is shown in Figure~\ref{fig:nonterminaltransf}. Because expanded function is a sequence of EXORS (reversible function) of variables or products of variables, the number of obtained gates is given by $e = 2^{n-1}$; where $n$ is the number of control bits.
\label{def:nonterminaltransf}
\end{corollary}
\begin{proof}
The proof is simple because the Non-Terminal Expansion concerns only product of single literal terms that are a sequence of XOR gates. As EXOR is a reversible logic gate, it always generates $2^n/2$ zeros and ones. Thus an EXOR of variables of the form $a\oplus b\oplus \dotso k\oplus l$ will generate an output function with exactly $2^l/2$ ones and $2^l/2 = 2^{l-1}$.
\end{proof}
\begin{definition}[Linearized Quantum Circuit]
A quantum circuit is called linearized if any gates defined on the exactly the same control qubits can be permuted without any change  to the output function on the output variables.  
\end{definition}

Thus Figure~\ref{fig:nonterminaltransf}b is an example of a Linearized Quantum Circuit because all the gates can be permuted without modifying the output function. Notice that in the Linearized Quantum Circuit the only remaining bit operations are either the control or the unitary transform applied to the target qubit(s); this means that the Linearized Quantum Circuit is functionally equivalent to Reed-Muller with different gates. This allows us to define the Permutation Equivalent Gate.

\begin{definition}[Permutation Equivalent Gate]
Two gates $g_1$ and $g_2$ in a linearized quantum circuit are called Permutation Equivalent (PE) if
\begin{enumerate}
\item they are defined on completely different set of variables. 
\item they use the same output bit as the target of the quantum operator.
\end{enumerate}
\label{cor:peg}
\end{definition}

\begin{figure}[bht]
\centering
\subfloat[]{\label{fig:toffg}$
\Qcircuit @C=1em @R=.9em {
     \lstick{{1}} & \ctrl{2} & \qw &\qw &\qw &\ctrl{1}&\qw&\qw&\ctrl{1}&\qw\\
     \lstick{{b}} & \qw & \qw &\ctrl{1}&\qw& \targ&\ctrl{1}&\qw&\targ&\qw\\
     \lstick{{c}} & \gate{V} & \qw &\gate{V} &\qw &\qw &\gate{V^\dagger}&\qw&\qw&\qw\\
}
$}
\subfloat[]{\label{fig:toffl}$
\Qcircuit @C=1em @R=.9em {
     \lstick{{a}} & \ctrl{2} & \qw &\qw &\qw &\ctrl{1}&\qw&\ctrlo{1}&\qw\\
     \lstick{{b}} & \qw & \qw &\ctrl{1}&\qw& \ctrlo{1}&\qw&\ctrl{1}&\qw\\
     \lstick{{c}} & \gate{V} & \qw &\gate{V} &\qw &\gate{V^\dagger}&\qw&\gate{V^\dagger}&\qw\\
}
$}
\subfloat[]{\label{fig:toffle}$
\Qcircuit @C=1em @R=.9em {
     \lstick{{a}} & \ctrl{2} & \qw &\ctrl{1}&\qw&\ctrlo{1}&\qw&\qw&\qw\\
     \lstick{{b}} & \qw & \qw & \ctrlo{1}&\qw&\ctrl{1}&\qw&\ctrl{1}&\qw\\
     \lstick{{c}} & \gate{V} & \qw &\gate{V^\dagger}&\qw&\gate{V^\dagger}&\qw&\gate{V}&\qw\\
}
$}
\caption{\label{eq:lineartoffoli}(a) Toffoli gate, (b) its linearized equivalent and (c) a linearized Toffoli gate with a permuted CV gate.}
\end{figure}
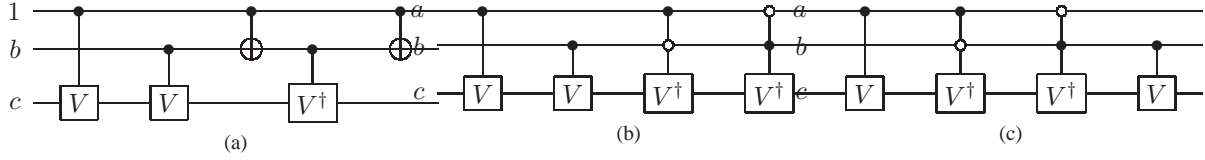

For instance, a Toffoli gate shown in Figure~\ref{fig:toffg} can be linearized to the Toffoli gate shown in Figure~\ref{fig:toffl} by factoring the $(a\oplus b)_b$ into two adjacent $a\bar{b}V^\dagger$ and $\bar{a}bV^\dagger$ quantum gates. Consequently, Figure~\ref{fig:toffle} shows that the CV gates (the right most gates in Figure~\ref{fig:toffg} and~\ref{fig:toffl} now can be moved arbitrarily in the resulting quantum circuit without changing the output function.

Note that the linearization created a new type of Toffoli gates: the multi-controlled V/V$^\dagger$ gates with mixed control (i.e. with positive/negative control inputs). Such gates (as $a\bar{b}V^\dagger$) might not always be realizable and thus we introduce the notion of \em Virtual Gates\em.

\begin{definition}[Virtual Gate]
A Quantum Virtual Gate is any quantum gate including such quantum gates that cannot be used for the circuit design but only as intermediary stages during minimization. All Quantum Virtual Gates must be transformed into real quantum gates once the minimization process is finished.
\end{definition}

\begin{corollary}[Creation of Virtual Gates]
The non-terminal gate transformation from Definition~\ref{def:nonterminaltransf} creates Virtual Gates with as many control input points as the expanded control bit function.
\end{corollary}
\begin{figure*}[bht]
\centering
\includegraphics[width=0.9\textwidth]{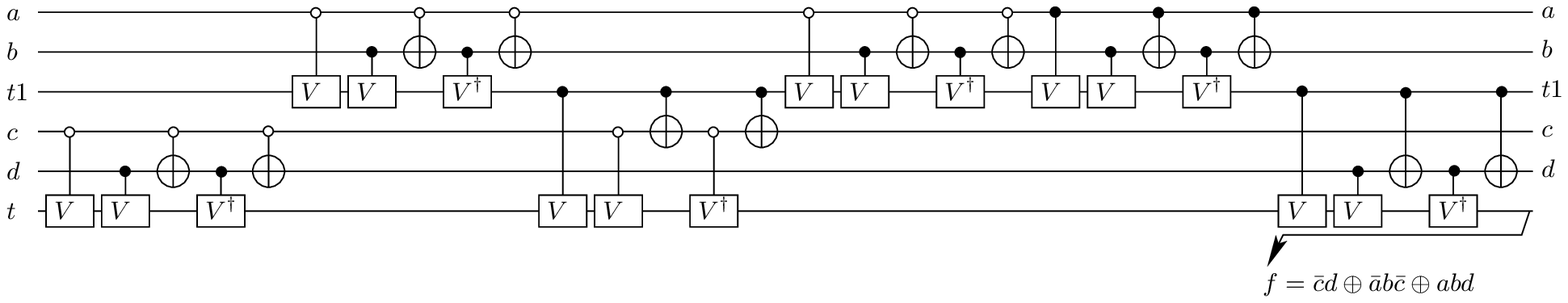}
\caption{\label{fig:secondexpansion}Circuit corresponding to the expansion described by eq.~\ref{eq:split2}}
\end{figure*}

For example, in a Toffoli gate, the expanded function during the linearization process is the $(a\oplus b)^V$ function and thus it will create 2 two-qubit controlled V gates $\bar{a}b^V$ and $a\bar{b}^V$. 

\begin{definition}[Quantum Operator Form (QOF)]
A QOF of a Boolean function is given by a set of quantum operator terms of a circuit in a linearized form. A QOF of a logic function contains only one type of a controlled function and is using Quantum Virtual Gates.
\end{definition}

\begin{example}[QOF of ESOP]
Let $f = \bar{c}d \oplus \bar{a}b\bar{c} \oplus abd$. Using the approach described above we can transform f into the QOF by consecutive steps. First, expand function f into a two-control-bit Toffoli gates and explicitly adding all intermediary variables denoted $tk$ and the output variable denoted by $t$:
\begin{equation}
\begin{split}
f = &\bar{c}d \oplus \bar{a}b\bar{c} \oplus abd\\
=&\bar{c}d_{t}\oplus [\bar{a}b_{t1}\circ t1c_t\circ \bar{a}b_{t1}] \oplus [ab_{t1}\circ t1d_t\circ ab_{t1}]
\end{split}
\label{eq:split1}
\end{equation}
The circuit corresponding to the expansion in eq.~\ref{eq:split1} is shown in Figure~\ref{fig:firstexpansion}. Notice that we remove the Toffoli gate restoring the intermediary ancilla bits as they are not necessary for further computation.
\begin{figure}[bht]
\centering
\includegraphics[width=0.4\textwidth]{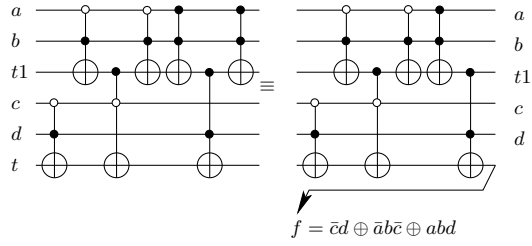}
\caption{\label{fig:firstexpansion}Circuit corresponding to the expansion described by eq.~\ref{eq:split1}}
\end{figure}
From eq.~\ref{eq:split1} we next convert all the Toffoli gates into the CV/CV$^\dagger$ and CNOT two-bit operators.

\begin{equation}
\begin{split}
f =&\bar{c}d_t\oplus \bar{a}b_{t1}\circ t1c_t\circ \bar{a}b_{t1} \oplus ab_{t1}\circ t1d_t\circ ab_{t1}\\
= &\bar{c}^V_td^V_t[\bar{c}\oplus d_d\circ d^{V\dagger}_t\circ \bar{c}\oplus d_d]\\
&\oplus \bar{a}_{t1}^Vb_{t1}^V[\bar{a}\oplus b_b\circ b^{V\dagger}_{t1}\circ\bar{a}\oplus b_b] \\
&\circ t1\bar{c}^V_t \circ \bar{a}_{t1}^Vb_{t1}^V[\bar{a}\oplus b_b\circ b^{V\dagger}_{t1}\circ\bar{a}\oplus b_b]\\
&\oplus a_{t1}^Vb_{t1}^V[a\oplus b_b\circ b^{V\dagger}_{t1}\circ a\oplus b]\circ t1d^{V\dagger}_t\\
=&\bar{c}^V_td^V_t[\bar{c}\oplus d_d\circ d^{V\dagger}_t\circ \bar{c}\oplus d_d]\\
&\oplus \bar{a}_{t1}^Vb_{t1}^V [\bar{a}\oplus b_b\circ b^{V\dagger}_{t1}\circ\bar{a}\oplus b_b]\\
&\circ t1^V_t\bar{c}_t^V [t1\oplus \bar{c}_c\circ c^{V\dagger}_t\circ  t1\oplus \bar{c}_c]\\
&\circ \bar{a}_{t1}^Vb_{t1}^V [\bar{a}\oplus b_b\circ b^{V\dagger}_{t1}\circ\bar{a}\oplus b_b]\\
&\oplus a_{t1}^Vb_{t1}^V[a\oplus b_b\circ b^{V\dagger}_{t1}\circ a\oplus b_b]\\
&\circ t1^V_td^V_t \circ [t1\oplus d_d\circ d^{V\dagger}_t\circ t1\oplus d_d]\\
\end{split}
\label{eq:split2}
\end{equation}

The circuit corresponding to eq.~\ref{eq:split2} is shown in Figure~\ref{fig:secondexpansion}. Observe that in eq.~\ref{eq:split2} some terms are in square brackets. These terms are the ones that will be expanded using the non-terminal gate expansion and will result in virtual quantum gates. This leads in eq.~\ref{eq:split3} that represents a non-minimized QOF:
\begin{equation}
\begin{split}
f = & \bar{c}^{\triangledown}_{t}d^\triangledown_{t}\bar{c}\bar{d}^{\triangledown\dagger}_{t} cd^{\triangledown\dagger}_{t}\\
&\oplus\bar{a}_{t1}^\triangledown b_{t1}^\triangledown \bar{a}\bar{b}^{\triangledown\dagger} _{t1} ab^{\triangledown\dagger} _{t1}\\
&\circ t1^{\triangledown}_t\bar{c}^\triangledown _{t}\bar{t1}\bar{c}^{\triangledown\dagger}_tt1c^{\triangledown\triangledown}_t\\
&\circ\bar{a}_{t1}^\triangledown b_{t1}^\triangledown \bar{a}\bar{b}^{\triangledown\dagger} _{t1} ab^{\triangledown\dagger} _{t1}\\
&\oplus a^{\triangledown}_{t1}b^{\triangledown}_{t1}\bar{a}b^{\triangledown\dagger}_{t1}a\bar{b}^{\triangledown\dagger}_{t1}\\
&\circ t1^{\triangledown}_td^{\triangledown}_t\bar{t1}d^{\triangledown\dagger}_tt1\bar{d}^{\triangledown\dagger}_t
\end{split}
\label{eq:split3}
\end{equation}

Notice that the V/V$^\dagger$ operators have been replaced by $\triangledown/\triangledown\dagger$ to indicate that some of the gates can be virtual gates and this is only a transitory form for circuit minimization. The corresponding circuit is shown in Figure~\ref{fig:expansion2}.
\begin{figure}[hbt]
	\centering
\includegraphics[width=0.5\textwidth]{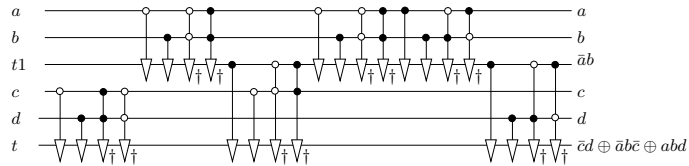}
\caption{\label{fig:expansion2} Circuit using the virtual gates for the function \em f\em}
\end{figure}
\end{example}

\section{Minimizing QOF}
\label{sec:minimization}

From eq.~\ref{eq:split3} several minimizations can be performed using the notion of the Permutation Equivalent Gates (Definition~\ref{cor:peg})) that are located on Uninterrupted Lines (Definition~\ref{def:uql}). The first step is to find uninterrupted lines. This can be simply done by finding such control lines that do not appear in any term's subscript. From eq.~\ref{eq:split3} variable C and D are uninterrupted and gates defined solely on these variables can be grouped together and combined. 

\begin{figure}[hbt]
	\centering
\subfloat[]{\label{fig:expansion2mina}\includegraphics[width=0.5\textwidth]{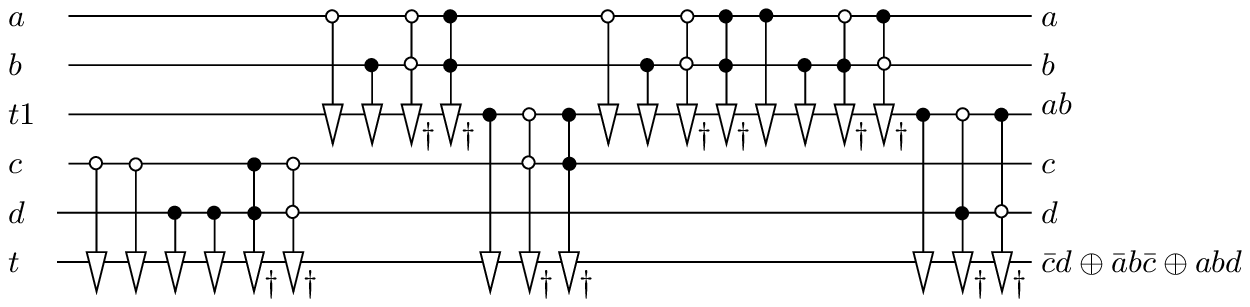}}\\
\subfloat[]{\label{fig:expansion2minb}\includegraphics[width=0.5\textwidth]{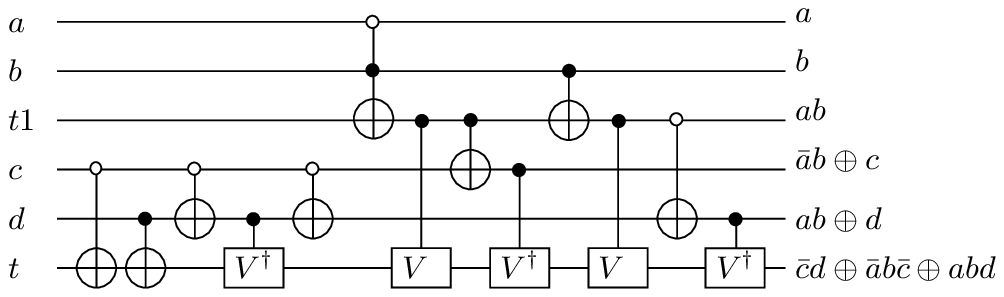}}\\
\subfloat[]{\label{fig:expansion2minc}\includegraphics[width=0.5\textwidth]{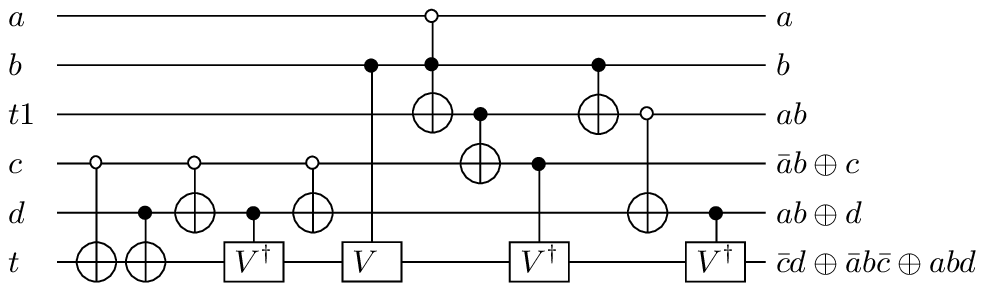}}
\caption{\label{fig:expansion2min} Steps of the minimization of the function \em f \em shown in Figure~\ref{fig:expansion2}}
\end{figure}

The minimizations steps are shown in Figure~\ref{fig:expansion2min}. Figure~\ref{fig:expansion2mina} shows that all gates that are defined on the same uninterrupted lines in Figure~\ref{fig:expansion2} are moved all to the right side of the circuit.

Next step is to look for gates that have the same output variable but are on interrupted line. The two topmost Toffoli gates are such gates and can be minimized by using the fact that it is possible to go from a two variable product term  to another with at maximum two EXOR forms~\cite{lukac:11}. This is shown in Figure~\ref{fig:expansion2minb}. It is also possible to perform more complex pattern matching transformations. Figure~\ref{fig:expansion2minc} shows the merging of $\bar{A}B_{t1}\circ t1_t^V$ and $\bar{A}B_{t1}\circ B^V_{t1}\circ t1_t^V$ to $B^V_t$.  In the QOF notation this amounts to search again for terms that are using the same control lines and then factoring out the interrupt points that allows to merge similar gates. Thus, from eq.~\ref{eq:split3} we obtain (we show only the concerned terms):
\begin{equation}
\begin{split}
f = & \bar{a}_{t1}^\triangledown b_{t1}^\triangledown \bar{a}\bar{b}^\triangledown _{t1} ab^\triangledown _{t1}\\
&\circ t1^{\triangledown}_t\bar{c}^\triangledown _{t}\bar{t1}\bar{c}^{\dagger}_tt1c^{\triangledown}_t\\
&\circ \bar{a}_{t1}^\triangledown b_{t1}^\triangledown \bar{a}\bar{b}^\triangledown _{t1} ab^\triangledown _{t1}\\
&\oplus a^{\triangledown}_{t1}b^{\triangledown}_{t1}\bar{a}b^{\triangledown\dagger}_{t1}a\bar{b}^{\triangledown\dagger}_{t1}\\
&\circ t1^{\triangledown}_td^{\triangledown}_t\bar{t1}d^{\triangledown\dagger}_tt1\bar{d}^{\triangledown\dagger}_t
\end{split}
\label{eq:split4}
\end{equation}
First, the two topmost Toffoli gates $\bar{a}_{t1}^\triangledown b_{t1}^\triangledown \bar{a}\bar{b}^\triangledown _{t1} ab^\triangledown _{t1}$ and $a^{\triangledown}_{t1}b^{\triangledown}_{t1}\bar{a}b^{\triangledown\dagger}_{t1}a\bar{b}^{\triangledown\dagger}_{t1}$ result in the second Toffoli gate being transformed to $b^{\triangledown}_{t1}b^{\triangledown}_{t1}$. 

\begin{equation}
\begin{split}
f = & \bar{a}_{t1}^\triangledown b_{t1}^\triangledown \bar{a}\bar{b}^\triangledown _{t1} ab^\triangledown _{t1}\\
&\circ t1^{\triangledown}_t\bar{c}^\triangledown _{t}\bar{t1}\bar{c}^{\dagger}_tt1c^{\triangledown}_t\\
&\oplus b^{\triangledown}_{t1}b^{\triangledown}_{t1}\\
&\circ t1^{\triangledown}_td^{\triangledown}_t\bar{t1}d^{\triangledown\dagger}_tt1\bar{d}^{\triangledown\dagger}_t
\end{split}
\label{eq:split5}
\end{equation}
Second, as already introduced above and using the same method as in the previous case - searching for gates defined on the same control variables: two CV gates defined $t1^{\triangledown}_t$ can be combined to:

\begin{equation}
\begin{split}
f = & \bar{a}_{t1}^\triangledown b_{t1}^\triangledown \bar{a}\bar{b}^\triangledown _{t1} ab^\triangledown _{t1}\\
&\oplus b^{\triangledown}_t \circ \bar{c}^\triangledown _{t}\bar{t1}\bar{c}^{\dagger}_tt1c^{\triangledown}_t\\
&\oplus b^{\triangledown}_{t1}b^{\triangledown}_{t1}\\
&\circ d^{\triangledown}_t\bar{t1}d^{\triangledown\dagger}_tt1\bar{d}^{\triangledown\dagger}_t
\end{split}
\label{eq:split6}
\end{equation}

Finally, the form can be re-composed into Toffoli gates and is shown in Figure~\ref{fig:expansion2minc}.

\section{Canonicity of QOF}
\label{sec:canonicity}
One of the important properties of PPRM is that it is a canonical form. Similarly, a desired property of the QOF is to be canonical. In the next section we provide a proof that under certain restrictions the QOF is a canonical representation and will can be called Quantum Operator Canonical Form (QOCF). 

\begin{definition}[Toffoli Gates Reduction]
Two Toffoli gates $T1$ and $T2$ defined on $m$ and $n$ control variables, respectively, can be reduced to one Toffoli gate with $m$ control bits and to one XOR gate and one Toffoli gate with $j$ control bits if:
\begin{equation}
\{v_0,\ldots,v_{j},\ldots,v_{n}\} = R(u_0,\ldots,u_j)P(v_{j+1},\ldots,v_m\}
\end{equation}
with $\{u_0,\ldots,u_m\}$ and $\{v_0,\ldots,v_n\}$ being the polarities of the control variables of the $T1$ and $T2$ gates, respectively. The $P$ term represents the product of $m-j+1$ variables and the $R$ term represents the following irreducible and non-expanding operation:
\begin{equation}
R= u_e\ldots u_{g}\oplus u_{e}\ldots u_{g+1}
\end{equation}
\label{def:reduction}
\end{definition}

\begin{figure}[bht]
\centering
\subfloat[\label{fig:reductionexamplea} $abde$]{
\includegraphics[width=0.13\textwidth]{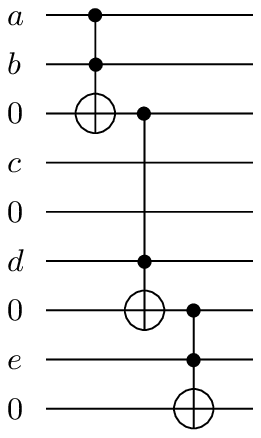}
}
\subfloat[\label{fig:reductionexampleb}$a\bar{b}de$]{
\includegraphics[width=0.13\textwidth]{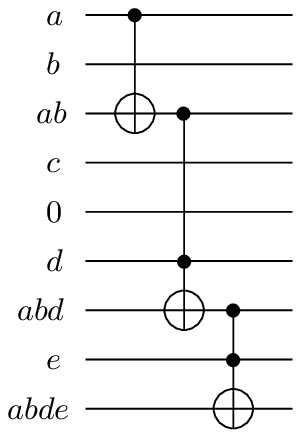}
}
\subfloat[\label{fig:reductionexamplec}$ab\bar{d}e$]{
\includegraphics[width=0.13\textwidth]{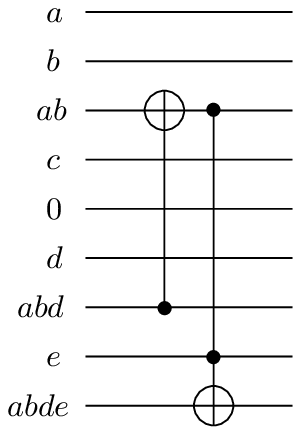}
}\\
\subfloat[\label{fig:reductionexampled}$abcde$]{
\includegraphics[width=0.13\textwidth]{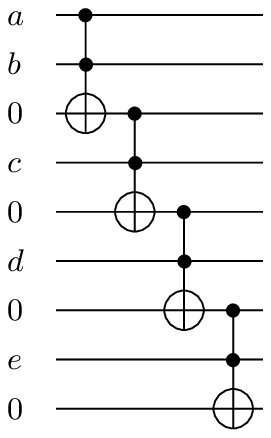}
}
\subfloat[\label{fig:reductionexamplee}$abc\bar{d}e$]{
\includegraphics[width=0.13\textwidth]{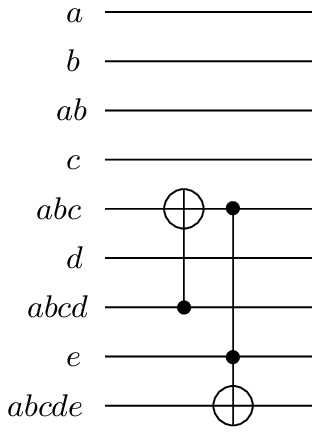}
}
\subfloat[\label{fig:reductionexamplef}$(abcd\oplus a)e$]{
\includegraphics[width=0.13\textwidth]{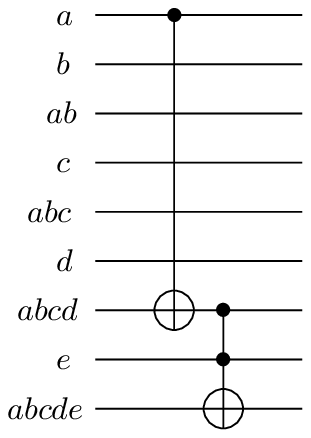}
}
\caption{\label{fig:reductionexample}Example of non reducible and reducible Toffoli gate reductions.}
\end{figure}

Examples of such reductions are shown in Figures~\ref{fig:reductionexamplea}-~\ref{fig:reductionexamplee}. Observe that this reduction means that any product of literals (a Toffoli gate) can be moved on the K-map to one of its adjacent minterms by the means of the transformation from Definition~\ref{def:reduction}. Figure~\ref{fig:reductionexamplea} is a term from which new terms are obtained by the transformations shown in Figures~\ref{fig:reductionexampleb} and~\ref{fig:reductionexamplec}. 

Similarly the term shown in Figure~\ref{fig:reductionexampled} is transformed into new terms using the circuits shown in Figures~\ref{fig:reductionexamplee} and ~\ref{fig:reductionexamplef}. Observe that the transformation shown in Figure~\ref{fig:reductionexamplef} is not a direct transformation as defined in Definition~\ref{def:reduction} and is shown as counter example.

\begin{theorem}[Weak Canonicity of QOF]
A QOF of a Reed-Muller after the application linearization minimization  (Section~\ref{sec:minimization}) and the transformation from Definition~\ref{def:reduction} is canonical with respect to the polarity of used literals if and only if variables wires are not altered and not permuted.
\label{th:canonic}
\end{theorem}
Observe that based on Theorem~\ref{th:canonic}, all intermediary and final products of literals can be created and manipulated only on the ancilla wires.
\begin{proof}
The remaining terms in the QOF are either defined on completely different variables (disjoint support set), on terms that are not adjacent on the K-map table or terms that are split by different Toffoli gates (Figure~\ref{fig:nonreducible}).
\begin{figure}[bht]
\centering
\subfloat[\label{fig:nonreduciblea}]{
\includegraphics[width=0.13\textwidth]{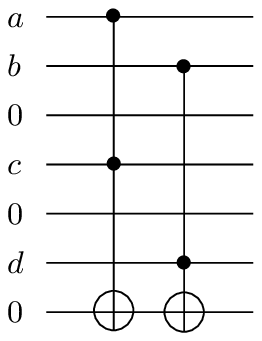}
}
\subfloat[\label{fig:nonreduciblea}]{
\includegraphics[width=0.13\textwidth]{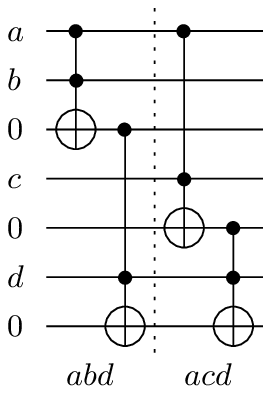}
}
\caption{\label{fig:nonreducible} Example of (a) disjoint support set  product terms  and (b) of terms that cannot be minimized without reordering the variables.}
\end{figure}
Figure~\ref{fig:nonreduciblea} shows two Toffoli gates that cannot be minimized because they are defined on completely variables - also called disjoint support set. Figure~\ref{fig:nonreduciblea} shows that terms $abd$ and $acd$ are both split by two Toffoli gates. Finally, Figure~\ref{fig:nonreduciblea} show that if the variables are reordered from $abcd$ to $adbc$ one Toffoli gate can be removed thus minimizing the circuit.

As was shown, the linearization allows to remove all compatible terms defined on same variables (even within other more complex Toffoli gates) and adjacent Toffoli gates can be minimized by CNOT gate replacement. Consequently, a minimization that would further reduce the circuit would also require variable change, additional ancilla bit or the change of the value of a variable wire. Example of such transformations are for instance template reduction~\cite{maslov:08,miller:10}.
\end{proof}

The algorithm to achieve the weak canonical form of the QOF can be obtained by combining an ordering algorithm such as the one described in~\cite{lukac:11a} with a minimization procedures of the linearized quantum circuits. An algorithm for this approach is shown below in the pseudo-code~\ref{alg:algo1}.

\begin{algorithm}
\caption{\label{alg:algo1}Pseudo-code for generating Weakly Canonical QOF.}
\begin{algorithmic}[1]
\STATE Order the XOR form of the circuit using Algorithm from ~\cite{lukac:11a}
\STATE Minimize the XOR form of the circuit using techniques from ~\cite{lukac:11a}
\STATE Minimize the resulting circuit using the linearization of the quantum circuit
\STATE Minimize the circuit using the Toffoli gate reduction method from def.~\ref{def:reduction}
\end{algorithmic}
\end{algorithm}

The importance of the weak canonicity is related to minimization and representation of quantum Boolean circuits. A canonical representation for quantum circuits is advantageous over the classical Reed-Muller because it not only it shows a non reducible representation using only truly quantum gates but it is also useful to be used as basis for other minimization methods for quantum circuits.

\section{Conclusion}
\label{sec:conclusion}

In this paper we presented an extension of the work on the symbolic operator approach introduced in~\cite{lukac:08}. The QOF can be used as a universal language to specifying a quantum function or a quantum circuit with a set of unitary operators and provides a set of tools to manipulate and minimize them. 

Future work includes the introduction of efficient rules for the minimization of such expressions and an algorithm for verification of results.

\section*{Acknowledgments}

P. Kerntopf was supported in part by the Polish Ministry of Science and Higher Education under Grant 4180/B/T02/2010/38.

\bibliographystyle{plain}

\bibliography{./main}

\end{document}